\documentclass[runningheads]{llncs}
\usepackage{graphicx}
\usepackage{url}
\usepackage{cite}
\usepackage{amsmath}
\usepackage{adjustbox,lipsum}
\usepackage{multirow}
\begin{document}
\title{QSOC: Quantum Service-Oriented Computing \thanks{European Commission grant  no.  825480  (H2020),  SODALITE.}}
%
%
\author{Indika Kumara\inst{1,2}\and
 Willem-Jan Van Den Heuvel\inst{1,2}\and
Damian A. Tamburri\inst{1,3}}
\authorrunning{I. Kumara et al.}
%
\institute{Jheronimus Academy of Data Science, Sint Janssingel 92, 5211 DA 's-Hertogenbosch, Netherlands
\and
Tilburg University, Warandelaan 2, 5037 AB Tilburg
\email{\{i.p.k.weerasinghadewage,w.j.a.m.vdnHeuvel\}@tilburguniversity.edu}
\and
Eindhoven University of Technology, 5612 AZ Eindhoven
\email{\{d.a.tamburri\}@tue.nl}}
\maketitle              
\begin{abstract}
Quantum computing is quickly turning from a promise to a reality, witnessing the launch of several cloud-based, general-purpose offerings, and IDEs. Unfortunately, however, existing solutions typically implicitly assume intimate knowledge about quantum computing concepts and operators. This paper introduces Quantum Service-Oriented Computing (QSOC), including a model-driven methodology to allow enterprise DevOps teams to compose, configure and operate enterprise applications without intimate knowledge on the underlying quantum infrastructure, advocating knowledge reuse, separation of concerns, resource optimization, and mixed quantum- \& conventional QSOC applications.

\keywords{Quantum Computing  \and Services \and SOA \and MDA \and Model-driven Engineering \and Reference Architecture.}
\end{abstract}

\section{Introduction}
Cloud services are being heralded as the key computing technology to unlock the massive computing power touted by quantum computing~\cite{FrankQuantumCloudcloser20}. Quantum computing holds the potential to boost memory exponentially larger than its apparent physical size, treat a exponentially large set of inputs in parallel, while marking a leap forward in computing capacity in close-to Hilbert space~\cite{Preskill2018quantumcomputingin}. 

Quantum computing seems to be of a specific value-add in industry verticals such as healthcare, financial services, and, supply chains~\cite{bova2021commercial}. For example, quantum computing may help to significantly fuel the speed - and harness effectiveness- of real-time detection of potential fraud and anomalies in astronomic amounts of financial transactions at a national or even global scale. For these reasons, according to \textit{ResearchAndMarkets.com}, the market for quantum computing is believed to rise as big as \$6 billion by 2023.

\subsection{The not (yet) delivered promise of Quantum Computing}
The basic fabric of quantum computing constitute qubits with fascinating yet "quirky" principles like entanglement and superposition- strongly interconnected, multi-state quantum particles in a quantum system. This allows them to not only simulate non-deterministic physical quantum systems, but also, crunch large numbers, run simulations and so on that have not been possible before with conventional compute infrastructures. 

However, this tantalizing promise of quantum computing is not delivered yet. 

Indeed, current quantum infrastructure suffer from various serious shortcoming~\cite{Preskill2018quantumcomputingin, QCOR}. Firstly, quantum algorithms are in practice only more efficient for certain classes of problems only, notably problems that can be only solved for bounded error, quantum, polynomial time (BQP). Secondly, quantum computers are (still) vulnerable for the lightest disturbances, causing decohere (loss of state). Thirdly, quantum applications embrace a single-application-single-model approach; implying they assume a one-to-one mapping to a stand-alone quantum compute infrastructure, making them notoriously hard to port.

By now, several (open source) SDKs that strongly draw on Python are offered by vendors such as IBM (QisKit), Microsofts Quantum Kit, to code, debug and test applications to develop quantum applications for quantum hardware. In this way, programmers may focus (more) on application-level problem-solving and coding while developing their quantum programs~\cite{QCOR}. In light of these developments large industry vendors have started to cloud-enable quantum computing infrastructures, witnessing launching Quantum Computing as a Service (QCaaS) models, such IBM's Quantum Experience, Microsoft's Azure Quantum, and, Amazon's Braket~\cite{FrankQuantumCloudcloser20}. Unfortunately however, these offerings typically require intimate understanding of the concepts and workings of quantum computing.

Another more recent trend commences to relax the single-application-single-model paradigm, departing from the stand-alone assumption of quantum applications. Hybrid application models have been suggested where quantum- and traditional computing infrastructures co-exist, allocating BQP parts of the application to noise quantum devices, and other less demanding parts to traditional compute infrastructures, reducing the risk of unstable, and unreliable applications~\cite{QCOR}. What is lacking to date is approaches that guide developers in this portioning process.

\subsection{SOA-enabling Quantum Computing for the Enterprise}
The above developments pose serious challenges to enterprise application developers to be able to effectively exploit the capacity of quantum compute hardware and the algorithms and SKDs tied to them, whilst at the other hand delivering industry-strength applications. We believe it is fair to assume that SOA designers and developers often simply lack the knowledge and capabilities to perform the low-level plumbing activities assumed by quantum compute infrastructures. They would have to go through a steep learning curve of pulses, circuits, Hamiltonian parameters, and complex algorithms. 

Workflow-oriented approaches such as \cite{FrankQWorkflow} certainly provide a key building block to enable the orchestration of classical applications and quantum circuits. Unfortunately, however, they are limited in there capacity to systematically reuse containerised and parameterized knowledge along a structured lifecycle model. In addition, they are typically application context agnostic, and do not explicitly take into consideration QoS- and resource constraints.

What they would require is methodologies and tools to design, develop, deploy, monitor and manage applications that may be (partially) executed on quantum infrastructure without much concerns of the underpinning details of quantum mechanics. At the same time, the enterprise application must be dynamically amendable and dependable, and thus able to be deal with (unexpected) changes in the underlying quantum hardware infrastructure. A very promising way for this seems to borrow proven concepts from model-driven engineering and SOA. In the rest of this vision paper we will explore further this.

\section{Quantum Service-Oriented Architecture}

\begin{figure}[!t]
\centering
  \includegraphics[width=0.8\textwidth]{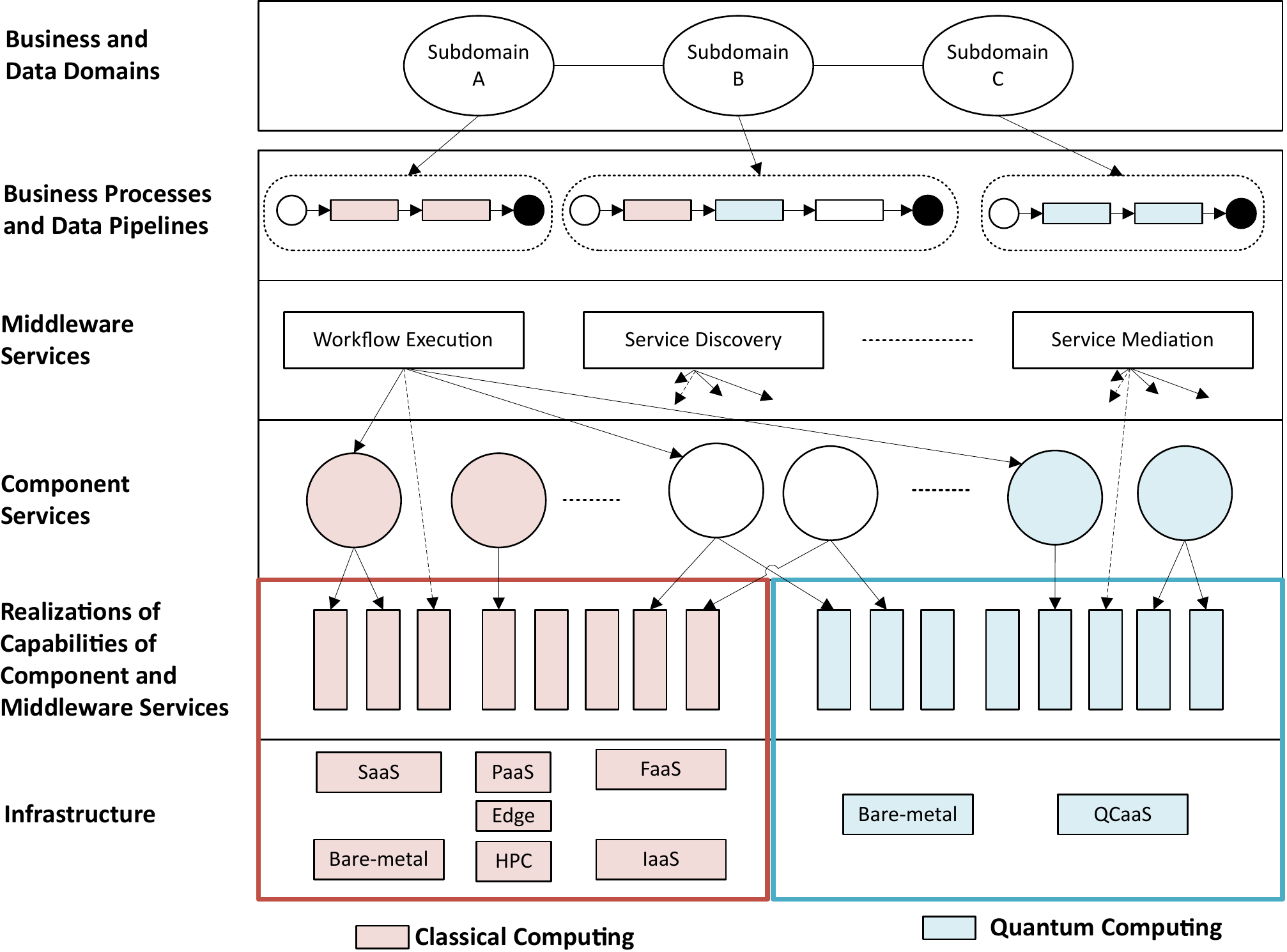}
\caption{Quantum Service-Oriented Architecture}
\label{qsocarch}      
\end{figure}

Inspired by the classical layered SOA~\cite{papazoglou2008web}, we have developed a layered reference architecture for QSOC, namely QSOA (Quantum Service-Oriented Architecture). QSOA provides an architectural framework for building applications exploiting a collection of reusable functional units (services) with well-defined interfaces and implemented using a well-balanced mix of classical and quantum (hybrid) computing approaches. Figure \ref{qsocarch} depicts the overall stratified QSDA architecture, which consists of the following six layers, starting from conceptual problem domain to the physical compute infrastructure: business or data domain, business processes or data pipelines, middleware services, component services, service realizations, and computing infrastructure. 

The topmost layer of QSOA captures the results of the problem domain analysis. To develop the QSOA applications (the technical solutions) that are in synergy with the business problem domain, we advocate the Domain Driven Design (DDD)~\cite{evans2004domain}. The DDD first partitions the problem domain into process- or data-oriented sub-domains. For example, we can split an online auction business domain into sub-domains such as auction, seller, listing, dispute resolution, and membership. When analyzing the problem domain, it is crucial to identify those segments of the business domains where the quantum computing adds value, for example, in terms of performance, cost, adding to its competitive advantage. 

Enterprise data can be partitioned into \textit{operational data} and \textit{analytical data}. Operational data is used to run the day-to-day operations of the business. Analytical data presents a temporal and aggregated view of the facts of the business over time, derived from operational data, and enables making data-driven business decisions. Typically this data is of special importance to decision makers at the tactic and strategic level. Business processes produce and consume operational data, whilst data pipelines, including ML pipelines and classical ETL (extract, transform, load) pipelines, typically ingest and generate analytical data. 

The DDD can be applied to segment the analytical data space into data domains aligned with the business domains (namely, the \textit{data mesh}~\cite{datamesh} approach). SOA and its recent manifestation \textit{microservices} enable building flexible and decentralized applications that support and automate business processes and data pipelines in enterprises. SOA applications seems a natural technology vehicle to leverage  quantum computing to perform computationally intensive tasks such as optimization, sampling, and machine learning~\cite{Holy}. 

In DDD, the concept in the solution domain that corresponds to a sub-domain constitutes the \textit{bounded context (BC)}, which is simply the boundary within a sub-domain where a particular domain model applies. Each BC includes its own domain model and discrete solution artifacts such as business processes, services, models, and requirements. A systematic approach to carving out solution artifacts from a BC is to apply tactical DDD patterns, especially for identifying microservices with natural, clean boundaries. Additional, "auxiliary"/"secondary" artifacts (e.g., speed-layer microservices) may be needed for achieving the non-functional requirements of the domain. The quantum computing can also be considered as a strategy for supporting non-functional requirements. 

Once the primary and secondary class microservices have been pinpointed and defined in clean, separate service definitions, the next step is to decide which microservices are to be implemented relying on classical computing or quantum computing, and/or, hybrid computing. Our QSOA assists in breaking down this decision process in a series of structured steps, promoting separation of concerns and loose coupling between the layers, and decision made in each of them.

As explained above, business processes encompass (primary) business services and (secondary) utility services such as services implementing calculations and data processing algorithms. Such utility services as well as the computationally intensive data processing services in the data pipelines can be suitable candidates to be implemented using the hybrid or quantum computing.  For instance, a drone delivery process may be instrumented with a real-time routing service to calculate the flight course for good deliveries. This service can be implemented using the quantum computing as computing the optimal route at real time is computationally expensive. 

Next, the middleware layer hosts, executes, monitors, and manages processes, pipelines and services. The implementation of middleware is likewise a good candidate to exploit quantum computing. Notably, computationally expensive capabilities such as service selection algorithms, service discovery algorithms, and machine learning based functions, seems ideal candidates. 

Following this line of reasoning and decision-making, there will emerge a set of business processes and data pipelines effectively implemented through synthesizing middleware and component services coded for, and deployed on classical, quantum or hybrid computing models.   

In this way, QSOC applications and middlewares are deployed over multiple, well-managed infrastructures including cloud, edge, HPC, and quantum (the bottom layer in the QSOA). Portability and adaptability will be crucial to support seamless and dynamic switching between different types of infrastructures as they or their usage contexts evolve. The appropriate mechanisms need be developed to support the efficient migration of application components across multiple infrastructures at runtime. 

\section{Model-Driven Quantum SOA Application Engineering Methodology}
\begin{figure}[!t]
\centering
  \includegraphics[width=0.85\textwidth]{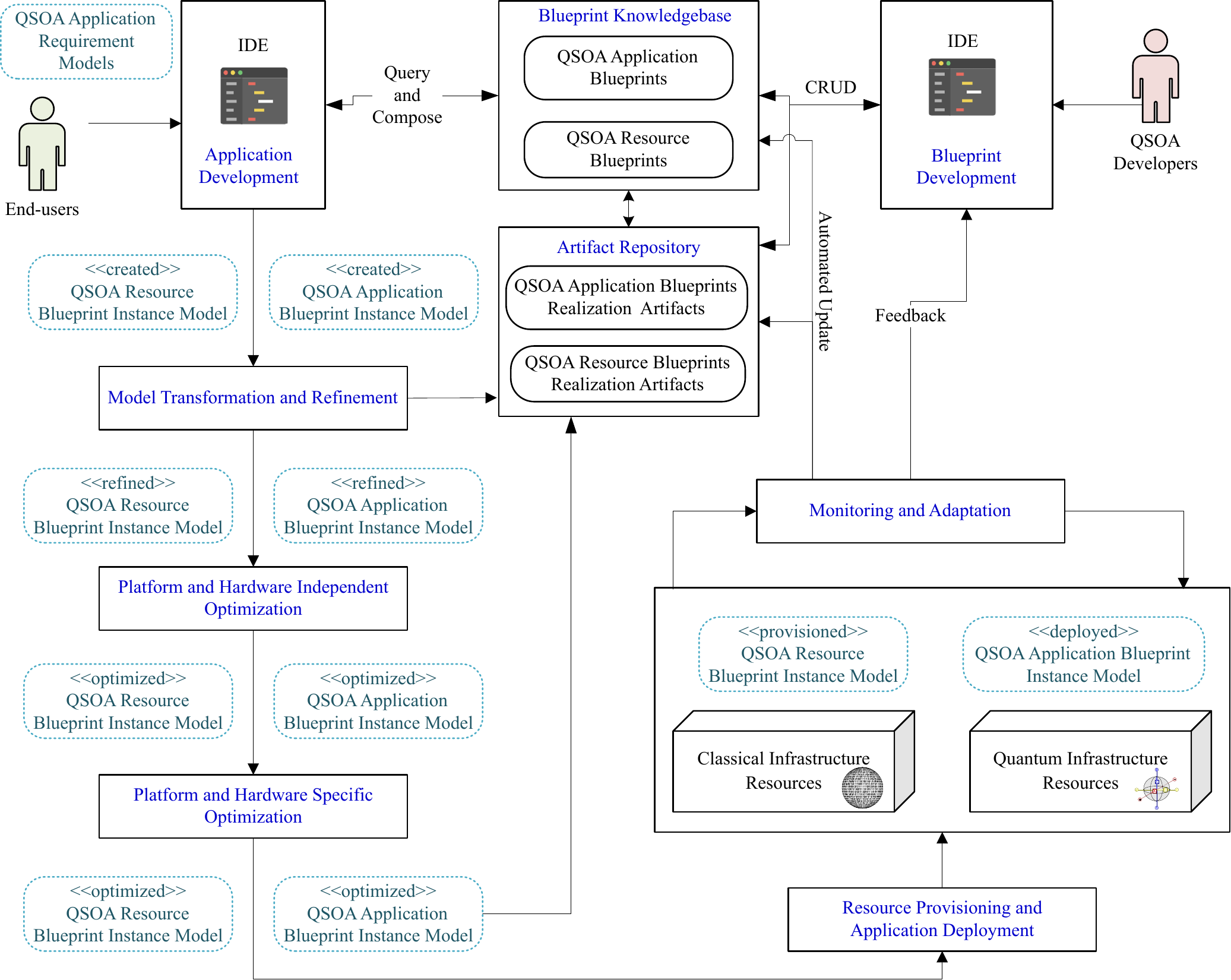}
\caption{Model-Driven QSOA Application Engineering Methodology}
\label{qsomethod}     
\end{figure}

In this section, we introduce a model-driven QSOA methodology (MD-QSOA) that facilitates the developers in designing, composing, deploying, executing, and managing the heterogeneous QSOA applications in a portable way, and to share and reuse the design and implementation artifacts, treating models as first class citizens. The methodology is designed to guide enterprise applications DevOps teams to implement and evolve the next generation of hybrid enterprise applications in a well-structured, repeatable and transparent manner.

Our approach is firmly grounded on model-driven engineering~\cite{MDA}, and has been inspired by our previous works on classical and cloud service engineering~\cite{papazoglou2006service, CloudBlue}. Notably, the QSOA methodology has benefited from ongoing works on heterogeneous application engineering~\cite{SODATLITE}. Figure~\ref{qsomethod} provides a graphical overview of the proposed methodology for engineering QSOA applications. 

The methodology has been designed on the assumption that the quantum applications are to be packaged as containerized microservices embracing the Quantum Algorithm/Application as a Service (QaaS) model~\cite{FrankQuantumCloudcloser20}. Moreover, we advocate the development of the QSOA artifacts (e.g., classical services, processes, and quantum applications) as \emph{customizable} artifacts~\cite{Tuan2} so that the end-users can dynamically assemble and (re)use the variants of the artifacts.    

The key element of our model-driven methodology constitutes reusable knowledge models, called \textit{blueprints}. We identify two main categories of blueprints: application blueprints and resource blueprints. The application blueprints are akin of computation independent model (CIM) and platform independent models semantically compartmentalizing meta-data on various aspect of a QSOA application, for example, requirements, service contracts, high-level process models, and deployment architecture models. The resource blueprints capture meta-data on operational-, performance-, and capacity requirements of infrastructure resources (e.g., cloud, edge, HPC, and quantum nodes) and the external services. 

\textit{Artifact Repository} stores versioned-controlled concrete realizations of blueprints, for example source codes for services and quantum algorithm codes, Docker files and images, and IaC (Infrastructure as code) scripts. \textit{Knowledgebase} consists of versioned-controlled ontologies or knowledge graphs of abstract blueprints to support discovery, matchmaking, reuse, composition, validation, and optimization of concrete blueprints. To implement the Knowledgebase, the existing ontologies can be adopted, for example, IaC ontologies~\cite{SODATLITE} and service ontologies~\cite{mcilraith2001semantic}. Ontologies enable representing both structural and semantics relationships of blueprints in an unambiguous manner, promoting reusability and interoperability. The existing automated ontology reasoning tools can be leveraged to implement the decision support required by the IDE and other components, for example, context-aware assistance of users at design-time (IDE), blueprint models enrichment taking into account domain knowledge (model transformers), and finding substitutable resources or services (the runtime adaptation support).

\textit{IDE} and \textit{Domain Specific Language (DSL)} are to support developers and end-users in their respective tasks. For example, the developers can model their infrastructure resources, the platforms, and the applications using blueprints, and store the models in the \textit{Knowledgebase}. They can add the implementations of algorithms and services into the \textit{Artifact Repository}, and store the metadata necessary to discover, validate and use the implementations. End-users can assemble and configure their QSOC application accurately and quickly making use of the context-aware intelligent content-assistance provided by the IDE powered by the \textit{Knowledgebase}. In this way, assisted mixing-and-maxing and optimization of requirements against resources allow faster and improved construction of application models by optimized selection and configuration of the resources and platforms to host the application components.   

The blueprint instance models produced by the IDE represent the platform independent knowledge models, for example, service contracts, business processes, policies, and deployment architecture models in the abstract. Such models need to be further refined and transformed into executable platform-specific models and code yet hardware-agnostic models/codes. For example, the deployment architecture models can be transformed into IaC programs such as TOSCA and Ansible~\cite{SODATLITE, TOSCA4QC}. The service variants can be generated from the service implementation artifacts retrieved from the \textit{Artifact Repository} using the service contracts. The model transformation and refinement process can utilize the reasoning capabilities of the \textit{Knowledgebase}. 

The refined blueprint instance models may not be optimal, and can have security vulnerabilities, performance anti-patterns, code smells and so on. Such issues should be detected and corrected. There exist a vast body of knowledge on quality assurance of program codes and software designs. Both classical and quantum algorithms can be optimized to achieve the best performance on a target environment~\cite{FrankQuantumSoftwareLifeCycle, SODATLITE}. In case of code artifacts, they can be built with the optimized compiler flags and libraries for a target environment, and the application parameters can also be auto-tuned~\cite{SODATLITE}. Domain (optimization) experts can use the \textit{Knowledgebase} to define the information about optimization options. The application optimizer can use such domain knowledge to select the correct options to map application configurations to the target hardware and build the optimized variants of the services and encapsulate as the containers.  

The refined, composed and optimized blueprint instance models for the application include the executable and deployable artifacts, for example deployment models as IaC scripts, services as containers, and processes as executable workflows. To deploy the applications, the IaC scripts can be executed using an orchestrator~\cite{TOSCA4QC, SODATLITE}. There exist different infrastructure providers, and they generally offer the REST APIs to create and management the resources in their infrastructures. These REST APIs hide the underling low-level resource orchestrators, and aid achieving interoperability of heterogeneous infrastructures. The orchestrator can be implemented as a meta-orchestrator that can coordinate low-level resource orchestrators using directly executing REST APIs or IaC scripts~\cite{SODATLITE}. 

The deployed application and the infrastructures can be continuously monitored by collecting various resource and application metrics to assure the satisfaction of the service level objectives (SLOs) for the application. The monitored data can also be used to determine sub-optimal use of resources, faults, anomalous behaviors, etc. This feedback loop allows for corrective, adaptive, and preemptive actions to be performed, for example, auto-scaling resources, selecting different service variants. In addition, this may involve moving application components from classical computers to cloud/HPC clusters to different types of quantum devices at runtime.

\section{Conclusion and Research Directions}
Quantum computing is now quickly becoming a reality, and the enterprises need to make sure their enterprise applications to be quantum-ready to offer unprecedented, warp-speed and non-linear services in a way that could not be offered before, keeping or gaining a competitive edge. 

In this vision paper, we coin the concept of quantum service-oriented computing (QSOC). In particular, we have underpinning and introduced a reference architecture for the quantum SOA model adopting the proven, industry-strength layered SOA model. In addition, we presented a novel model-driven methodology for realizing QSOA applications that makes it easier for enterprises to build portable and adaptive QSOA applications. 

Indeed, the results presented in this article are core research in nature. The paper is meant as a foundation for manifold of future research with a high academic and industrial relevance and impact. We propose three research directions for further exploring and realizing the vision of QSOC and MD-QSOC.

Firstly, we intend to further flesh out each layer of the QSOA model, improving our understanding on which issues are to be best solved by classical, quantum, and/or hybrid algorithms. Indeed, empirical studies as well as action-design methodologies for mapping computing models to problems considering multiple criteria are necessary. Moreover, the migration of classical and legacy algorithms/services to quantum and hybrid models also needs to be investigated as a logical continuation of our earlier works in SOC. 

Secondly, we plan to develop and validate the concepts, theories and techniques underpinning the model-driven methodology for engineering and managing QSOA applications. The blueprint models, customizable QSOA artifacts, IDE and DSLs, model transformation and refinement processes, semantic knowledgebase and decision support, platform/hardware independent/specific optimizations, orchestration, monitoring, and run-time adaptation all will have numerous research challenges. Understanding and resolving such research challenges would be crucial for successful longer-term adoption of quantum computing within enterprises to support their business processes and analytical data pipelines. 

Thirdly, and lastly, we intend to setup short-cyclic, "learning-by-doing" industrial experimentation with the MD-QSOA and the tools developed in the first- and second research lines, deriving patterns, improving the decision-model underpinning the methodology, and, improving our understanding and theoretical implications of the QSOA applications.

\bibliographystyle{splncs04}
\bibliography{main}

\end{document}